\documentclass[twocolumn]{pasj00}

\begin{document}
\SetRunningHead{Nagai et al. }{Physical Conditions of Molecular Gas in the Galactic Center}
\Received{2006/0x/xx}
\Accepted{2006/0x/xx}

\title{Physical Conditions of Molecular Gas in the Galactic Center}


%
\author{%
  Makoto \textsc{Nagai}\altaffilmark{1}
  Kunihiko \textsc{Tanaka}\altaffilmark{2}
  Kazuhisa \textsc{Kamegai}\altaffilmark{2}
  and
  Tomoharu \textsc{Oka}\altaffilmark{1}}
\altaffiltext{1}{Research Center for the Early Universe and Department of Physics\\
The University of Tokyo, 7-3-1 Hongo, Bunkyo-ku, Tokyo 113-0033}
\email{nagai@taurus.phys.s.u-tokyo.ac.jp}
\altaffiltext{2}{Institute of Astronomy, University of Tokyo, 2-21-1 Osawa, Mitaka, Tokyo
181-0015, Japan.}

\KeyWords{galaxies: nuclei---Galaxy: center---ISM: clouds---ISM: molecules} 

\maketitle

\begin{abstract}
We estimated physical conditions of molecular gas in the central molecular zone (CMZ) of the Galaxy, using our CO {\it J}=3--2 data obtained with the Atacama Submillimeter Telescope Experiment (ASTE) in conjunction with {\it J}=1--0 $^{12}$CO and $^{13}$CO data previously observed with the NRO 45m telescope.  The large velocity gradient (LVG) approximation was employed.  Distributions of gas density, kinetic temperature, and CO column density are derived as functions of position and velocity for the entire coverage of the CO {\it J}=3--2 data.  We fairly determined physical conditions for 69\%\
 of data points  in the CMZ with $\geq 1 \sigma$ CO detections. 
 Kinetic temperature was found to be roughly uniform in the CMZ, while gas density is higher in the 120-pc star forming ring than in the outer dust lanes.  Physical conditions of high {\it J}=3--2/{\it J}=1--0 features are also discussed.  

\end{abstract}

\section{Introduction}

Molecular gas concentrates in the inner $r\leq 200$ pc of the Galaxy, the Central Molecular Zone (CMZ).  The average density and kinetic temperature are much higher in the CMZ than anywhere else in the Galaxy (e.g., Morris \&\ Serabyn 1996).  The distribution and kinematics of molecular gas there is highly complex.  Its small-scale structure is characterized by arcs/shells, filaments, and a population of high-velocity compact clouds (HVCCs), suggesting a number of supernova explosions may be responsible for the boisterous gas kinematics and peculiar physical conditions there.  

We are performing a large-scale, high-resolution CO {\it J}=3--2 survey of the CMZ with the Atacama Submillimeter Telescope Experiment (ASTE) to pick up shocked gas in the CMZ completely (Oka et al. 2006).  As expected, we have detected many high {\it J}=3--2/{\it J}=1--0 features in the $\Delta l\times \Delta b\!=\!2\arcdeg\times 0.5\arcdeg$ area of the CMZ.  To understand the nature of these high ratio features, it is crucial to know physical conditions in them.  This paper presents our attempts to estimate physical conditions of molecular gas using our CO {\it J}=3--2 data in conjunction with {\it J}=1--0 $^{12}$CO and $^{13}$CO data previously obtained with the NRO 45m telescope (Oka et al. 1998).  The distance to the Galactic Center is assumed to be 8.5 kpc. 

\section{The Data}
We used three data sets of CO rotational lines. 
The first data set is the CO $J$=3--2 line obtained recently with the ASTE \citep{paper1}. 
The others are the CO $J$=1--0 and $^{13}$CO $J$=1--0 data obtained with the Nobeyama Radio Observatory 45 m telescope \citep{1998ApJS..118..455O}. These data sets cover a common region $-0.60$\arcdeg\ $\leq l \leq +1.50$\arcdeg , $-0.23$\arcdeg $\leq b \leq +0.23$\arcdeg . The velocity coverage is from $V_{\rm LSR}= -200$ km s$^{-1}$ to 200 km s$^{-1}$.  These data were regrided onto a 34\arcsec $ \times $34\arcsec $ \times $2 km s$^{-1}$ grid, and smoothed to a 1\arcmin $ \times $1\arcmin $ \times $2 km s$^{-1}$ (FWHM) with a Gaussian filter function. Resultant 1 $\sigma$ noise levels in the $T_{\rm mb}$ scale are summarized in Table \ref{tab:lines}. Total number of data points in this grid is 2284800. 

\begin{table}[hbt]
  \caption{Data sets}\label{tab:lines}
  \begin{center}\begin{tabular}{llll}
  \hline \hline
  Line & Telescope & Noise level & Reference \\
    \hline
  \hline
  CO $J$=3--2 & ASTE & 0.43 K & \citet{paper1} \\
  CO $J$=1--0 & NRO 45 m & 0.32 K & \citet{1998ApJS..118..455O} \\
  $^{13}$CO  $J$=1--0 & NRO 45 m & 0.28 K & \citet{1998ApJS..118..455O} \\
  \hline
  \end{tabular}\end{center}
\end{table}

\section{Analysis}
The large velocity gradient (LVG) model \citep{gol,sco} is employed to calculate intensities of line emission from a small volume within a molecular cloud. We calculated CO line intensities as functions of CO column density per unit velocity width $N_{\rm CO}$/d$V$, kinetic temperature $T_{\rm k}$, and number density of molecular hydrogen $n$(H$_2$) (hereafter we write it $n$). 
We developed a numerical method to derive the physical conditions ($N_{\rm CO}$/d$V$, $T_{\rm k}$, $n$) from given line intensities of CO $J$=1--0, $^{13}$CO  $J$=1--0, and CO $J$=3--2 emission ($T_{1-0}$, $T_{13}$, $T_{3-2}$). This calculation is a least square fit of the LVG model to the observed intensities. Here we describe our model, the least square fit, and its application to the data.

\subsection{LVG model}
Our LVG model uses the escape probability of a spherical cloud. 
The rotational levels of CO up to $J$=40 are considered. 
The model uses the collisional rate coefficients for de-excitation calculated by \citet{2001JPhB...34.2731F} and extrapolated by \citet{2005A&A...432..369S}. 
Collision partners are H$_2$ molecules, and the ortho/para ratio was assumed to be 3. 
The $^{12}$CO/$^{13}$CO abundance ratio is taken to be 24 \citep{1990ApJ...357..477L}. 
We did not make any explicit assumption on velocity gradient, CO/H$_2$ abundance, or depth of emitting region along the line of sight. 

The equations of statistical equilibrium are solved iteratively. Physical conditions on a grid which covers a domain $D$, 10$^{14}$ cm$^{-2}$(km s$^{-1}$)$^{-1} \leq N_{\rm CO}/{\rm d}V \leq 10^{19}$ cm$^{-2}$(km s$^{-1}$)$^{-1}$, 5 K $ \leq T_{\rm k} \leq $ 120 K, and 10$^{1}$ cm$^{-3}$ $ \leq n \leq $ 10$^{7}$ cm$^{-3}$. This parameter range can cover whole observed line intensities above the detection limit. 
The grid spacings are 0.1 dex for $N_{\rm CO}/{\rm d}V$ and $n$, 1 K for $T_{\rm k}$. 
For off-grid points, we analytically interpolate the line intensities on the grid, using a tricubic interpolation (\cite{2005IJNME..63....455L} for general discussions), to reduce the computational cost. 

\begin{figure*}[t]
  \begin{center}
    \FigureFile(120mm,80mm){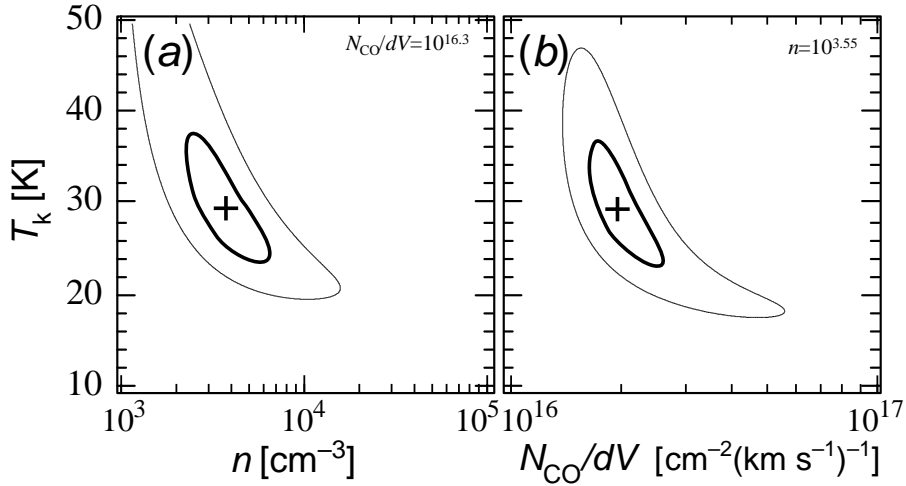}
  \end{center}
  \caption{Function $\chi^2(x)+f_{\rm cons}(x)$ for $T_{1-0}$=13 K, $T_{13}$=1 K, and $T_{3-2}$=10 K, ({\it a}) at $N_{\rm CO}/{\rm d}V$=10$^{16.3}$ cm$^{-2}$(km s$^{-1}$)$^{-1}$, ({\it b}) at $n$=10$^{3.55}$ cm$^{-3}$. Contours are 11.35 and 50. The cross indicates the minimum of the function. }
\label{fig:lvg}
\end{figure*}

\subsection{Fitting Procedure}
The problem we have to solve is to determine $x$=($N_{\rm CO}$/d$V$, $T_{\rm k}$, $n$) for given $y$=$(T_{1-0}, T_{13}, T_{3-2})$, whose elements are observed line intensities of CO $J$=1--0, $^{13}$CO  $J$=1--0, and CO $J$=3--2 respectively. 
For this purpose, we employed the least-square fit method. The best fit parameter $x$ gives a local minimum of the chi-square function 
\begin{equation}
\chi^2 (x)=\sum_{i} \{ y_i-T_i^{*}(x) \}^2/\sigma_i^2, 
\label{chi2}
\end{equation}
where $T_i^{*}(x)$ is the line intensity predicted by the model, and $\sigma_i$ is 1 $\sigma$ noise level of $y_i$. Note that the function is known as chi-square distribution with 3 degrees of freedom. The chi-square function become zero when the given line intensities $y$ can be reproduced by the model. 
We can assess the quality of fit $x$ by the value of $\chi^2$.

Line intensities converge to $ \sim T_{\rm k}$ in the optically thick thermalized region. In this region, it is difficult to constrain $n$ and $N_{\rm CO}$/d$V$, since line intensities stay nearly constant over wide range of parameters. 
A simple minimization of $\chi^2$  often returns unphysically high $n$ and $N_{\rm CO}$/d$V$ in the thermalized region. 
To suppress this divergence, we introduced a constraint function $f_{\rm cons}(x)$ and searched the minimum of $\chi^2(x)+f_{\rm cons}(x)$. The constraint function is written as 
\begin{eqnarray}
f_{\rm cons}(x)=\left\{ \begin{array}{ll}
C_{\rm cons} & (x \notin D) \\
0 & (x \in D, n \leq 0.5 n^{3-2}_{\rm crit}) \\
a \left(\log \frac{n}{0.5 n^{3-2}_{\rm crit}}\right)^2 & (x \in D, n > 0.5 n^{3-2}_{\rm crit}) \\
\end{array} \right.
\end{eqnarray}
where $C_{\rm cons}$ is an arbitrary large constant to restrict the solution to domain $D$, $n^{3-2}_{\rm crit}$ is the critical density of the CO $J$=3--2 transition (10$^{3.8}$ cm$^{-3}$), and $a$ is a positive constant which determines the strength of modification to suppress unrealistic high density. The form of $f_{\rm cons}(x)$ on $D$ is chosen so that it is 0 in $ n \leq 0.5 n^{3-2}_{\rm crit}$, quadratic in $ n > 0.5 n^{3-2}_{\rm crit}$, differentiable on $D$, and $a$ is taken so that $f_{\rm cons}=1$ at $n=n^{3-2}_{\rm crit}$. 
Figure \ref{fig:lvg} shows an example of $\chi^2(x)+f_{\rm cons}(x)$. 

To solve the minimization problem, we tested the steepest descendent method and the quasi-Newton method BFGS algorithm. Although they gave the same result in most cases, the latter is less stable and sometimes failed to get a global minimum. Thus we employed the former method in our calculations. 
We used a fixed initial value $N_{\rm CO}/{\rm d}V$=10$^{16}$ cm$^{-2}$(km s$^{-1}$)$^{-1}$, $T_{\rm k}$=25 K, and  $n$=10$^{3}$ cm$^{-3}$, for all parameter range. 
A local minimum of $\chi^2(x)+f_{\rm cons}(x)$ was obtained in any case. We used $\chi^2$ to evaluate the propriety of solutions. If $\chi^2 \leq 11.35$, which corresponds to 99 \% confidence of the chi-square distribution with 3 degrees of freedom, 
we can say that the parameters are well-determined.

\subsection{Existence Examination of Solution}

\begin{figure*}[t]
  \begin{center}
    \FigureFile(120mm,80mm){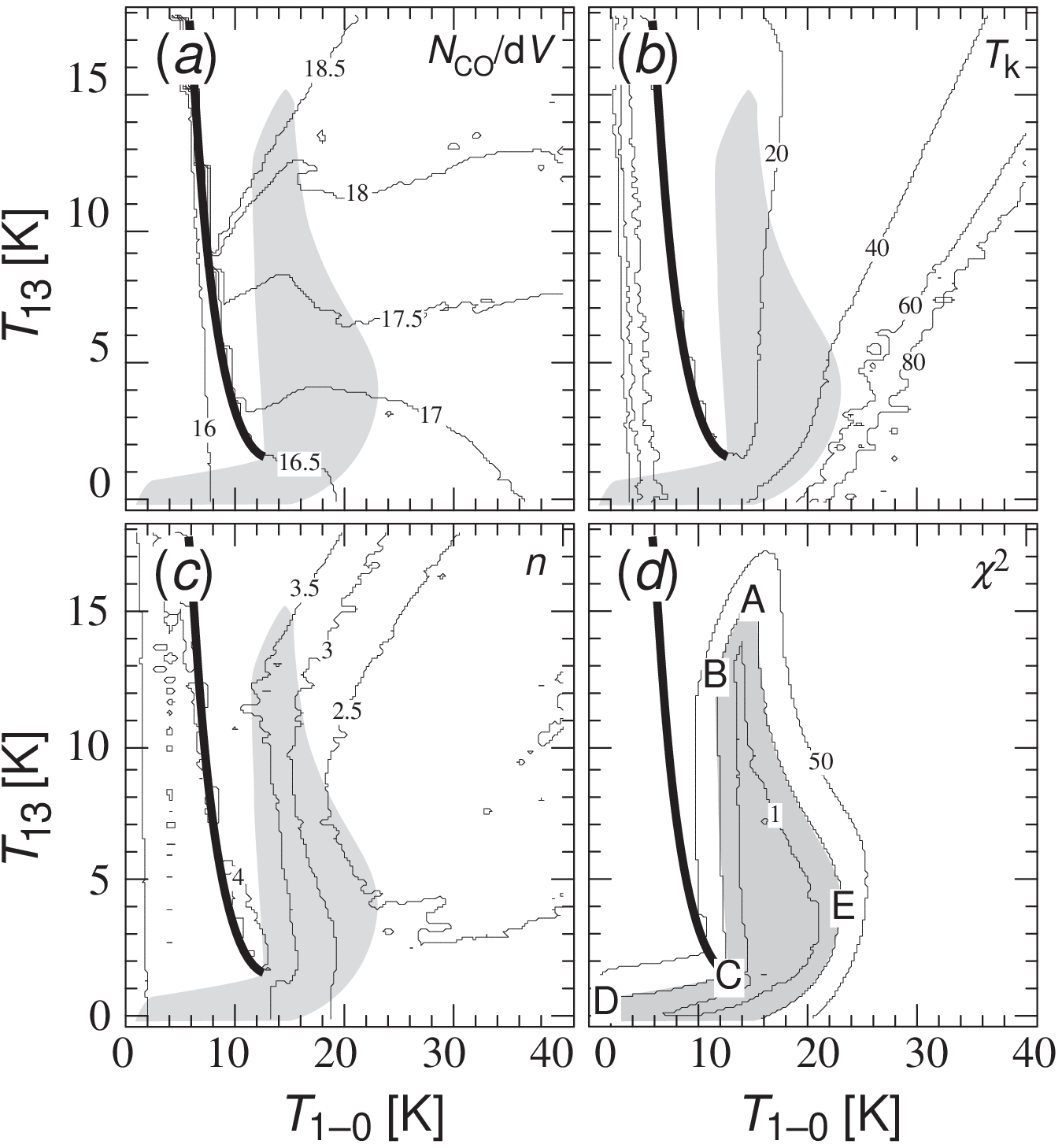}
  \end{center}
  \caption{Physical conditions determined by the LVG model. The CO $J$=3--2 line intensity is fixed to 10 K: ({\it a}) CO column density per unit velocity width, ({\it b}) kinetic temperature, ({\it c}) number density of molecular hydrogen, and ({\it d}) chi-square function. In ({\it a}) and ({\it c}) contours are labeled by logarithm to base 10. 
  
  A region where chi-square is less than 11.35 is shown in gray. Thick lines indicate the discontinuity line of best-fit parameters. }\label{fig:lvgi}
\end{figure*}

Figure \ref{fig:lvgi} shows results of the least-square fit for $T_{3-2}=10$ K and various $T_{1-0}$ and $T_{13}$. 
Parameters are well-determined in a region shown in gray. 
Vertexes on the boundary of the well-determined region are named A, B, C, and D. 
Each part of the boundary corresponds to a limiting case. 
Segment AB corresponds to the optically-thick limit where $T_{1-0}=T_{^{13}{\rm CO}}$. 
Segments BC and CD correspond to the local thermodynamical equilibrium (LTE) case for optically thick and thin limits respectively. This boundary comes from a condition that excitation temperature of $J$=1--0 is higher than that of $J$=3--2. 
Segments DE and EA correspond to the subthermal limit for optically thin and thick respectively. 
Kinetic temperature increases as the excitation temperature of $J$=3--2 becomes subthermal when lines are optically thin, and density decreases when optically thick. 
Out of the well-determined region, there is a line of discontinuity in parameters, indicated by a bold line in Figure \ref{fig:lvgi}. 

\subsection{Application to the Data}
We applied the fitting method for each pixel of the data described in \S 2. 
We assume that beam filling factors ($\eta_{1-0}$, $\eta_{13}$, $\eta_{3-2}$) are 1 for all lines. The smaller beam filling factors, $ \eta \leq 1$, generally give a higher temperature for most data points. In case of $\eta_{13} \leq \eta_{1-0} = \eta_{3-2}$, which often occurs in giant molecular clouds, kinetic temperature decreases while density increases slightly. 

Another breakdown of one-zone assumption occurs in a highly optically thick cloud.  In such a cloud, highly opaque $^{12}$CO line is severly self-absorbed, tracing the low-density, subthermally excited envelope, while less opaque $^{13}$CO line traces the high-density cloud core.  This situation gives $T_{13}$ similar to $T_{12}$, thereby one-zone LVG analyses give low-temperature, high-density solutions.  Unfortunately, $ \chi ^2$ of such improper solutions are not necessarily high.  A discontinuity of physical conditions appears at the boundary between the proper and improper solution areas in the {\it l-b-V} cube.   These improper solutions can be rejected by the behavior of $ \chi^2$ around the minimum; $ \chi^2$ around improper solutions are more sensitive to the temperature than those around proper solutions.  We employed the rejection criterion, $ \partial^2 \chi^2/\partial T_{\rm k}^2 > 10$.

The LTE assumption may hold well in such thermalized optically thick cloud cores.  In this case, $T_{13}$ can be related to $N_{\rm CO}$ by, 
\begin{eqnarray}
\int T_{13} {\rm d}V = (f_{12/13}+1)^{-1} \frac{1}{Z} \exp \left( -\frac{h \nu}{k T_{\rm ex}} \right) \nonumber \\ \times \frac{hc^3}{8 \pi k}\frac{3A_{10}}{\nu^2} \left( 1-\frac{I_{\nu }(T_{\rm bg})}{I_{\nu }(T_{\rm ex})} \right) \frac{1-e^{-\tau } }{\tau } N_{\rm CO}\\  \nonumber 
T_{13} = \{ I_{\nu }(T_{\rm ex})-I_{\nu }(T_{\rm bg}) \}(1-e^{-\tau })
\label{eq:column}
\end{eqnarray}
where $f_{12/13}$ is the abundance ratio $^{12}$CO/$^{13}$CO, $Z$ is the partition function,  $ \nu $ is the frequency, $T_{\rm ex}$ is the excitation temperature, $A_{10}$ is the Einstein A-coefficient, $I_{\nu }(T)$ is the Planck function, $T_{\rm bg}$ is background radiation temperature, and $ \tau $ is the line optical depth. 
Reducing eq.(3), we obtain 
\begin{eqnarray}N_{\rm CO}/{\rm d}V = 1.15 \times 10^{15} \left( T_{\rm ex} +2.64 \exp \frac{5.29}{ T_{\rm ex}} \right) \nonumber \\ \times \left(1+\frac{0.87}{ T_{\rm ex}}\right) \cdot \left(1-\frac{1}{2}\frac{ T_{13} }{ T_{\rm ex} }\right) T_{13} .\label{eq:columnapp}\end{eqnarray}
This means that $N_{\rm CO}/dV$ is approximately proportional to $T_{13}$.  If $T_{\rm ex}$ varies from 20 K to 50 K, $N_{\rm CO}/dV$ increases $\sim 70$ \%.  The discrepancy within a factor of two may not be serious for $N_{\rm CO}/dV$, considering it ranges over three orders of magnitude.

\section{Results and Discussion}
\begin{figure*}[btp]
  \begin{center}
    \FigureFile(120mm,80mm){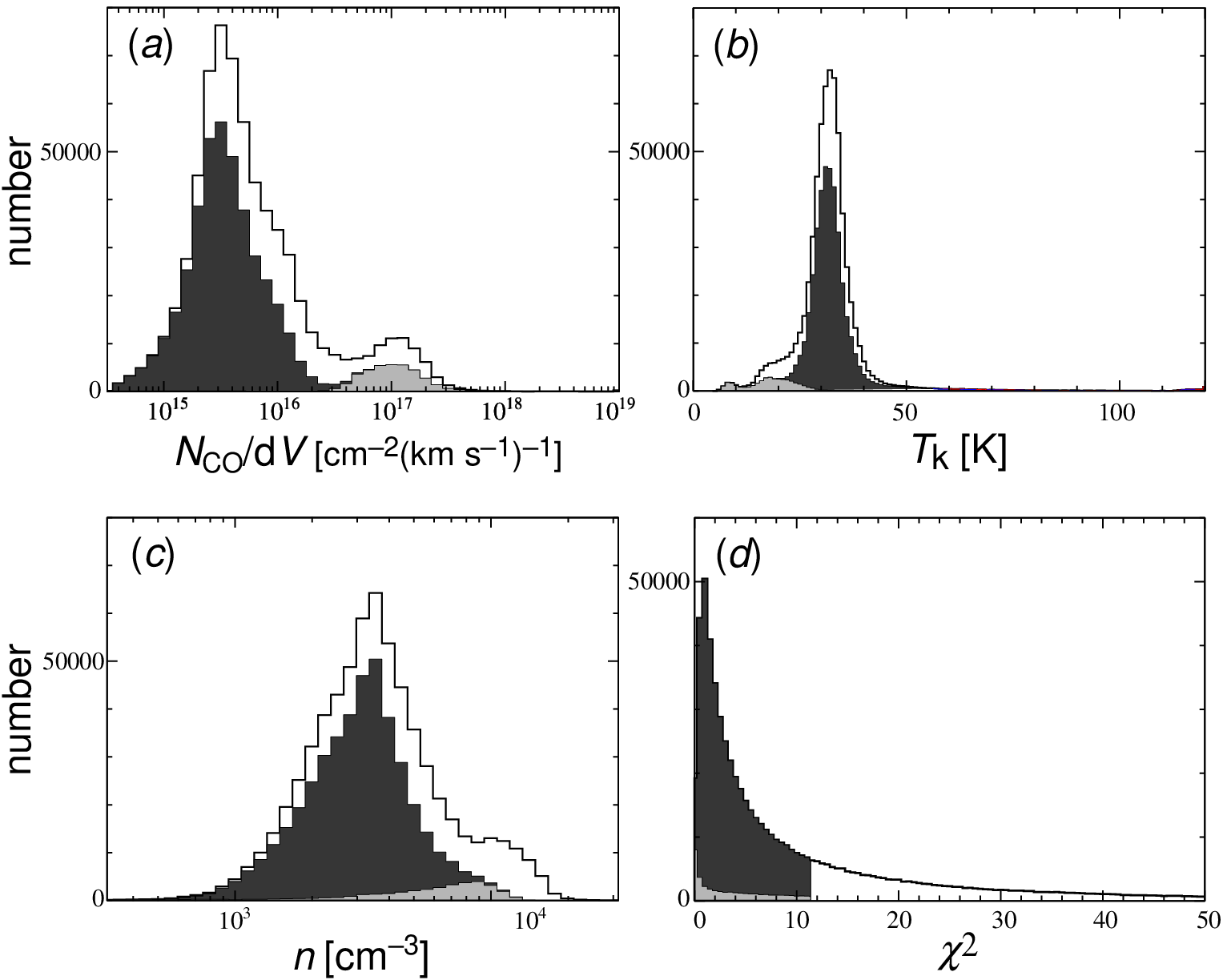}
  \end{center}
  \caption{Frequency distribution of the physical conditions: ($a$) CO column density per unit velocity dispersion interval, ($b$) kinetic temperature, ($c$) number density of molecular hydrogen, and ($d$) chi-square function. Improper solutions with chi-square less than 11.35 are shown in light gray. Proper solutions with chi-square less than 11.35 are shown in dark gray. }\label{fig:histgram}
\end{figure*}

We successfully obtained proper solutions with significant confidence ($\chi^2 \leq 11.35$) for 400841 data points out of 653550 points. Histograms of physical conditions and $\chi^2$ are shown in Figure \ref{fig:histgram}. 
The lower limit of $N_{\rm CO}/{\rm d}V$ is set by detection limits of spectral lines. $N_{\rm CO}/{\rm d}V$ reaches 10$^{19}$ cm$^{-2}$(km s$^{-1}$)$^{-1}$  only for the foreground arms where line optical depths are quite large, where the parameters contain large uncertainties. We had low kinetic temperatures ($ \sim $5 K) in the foreground arms. 
The highest temperature ($ \sim $120 K) was found in the vicinity of Sgr A. 

\subsection{Distribution of physical conditions}
\begin{figure*}[btp]
  \begin{center}
    \FigureFile(160mm,80mm){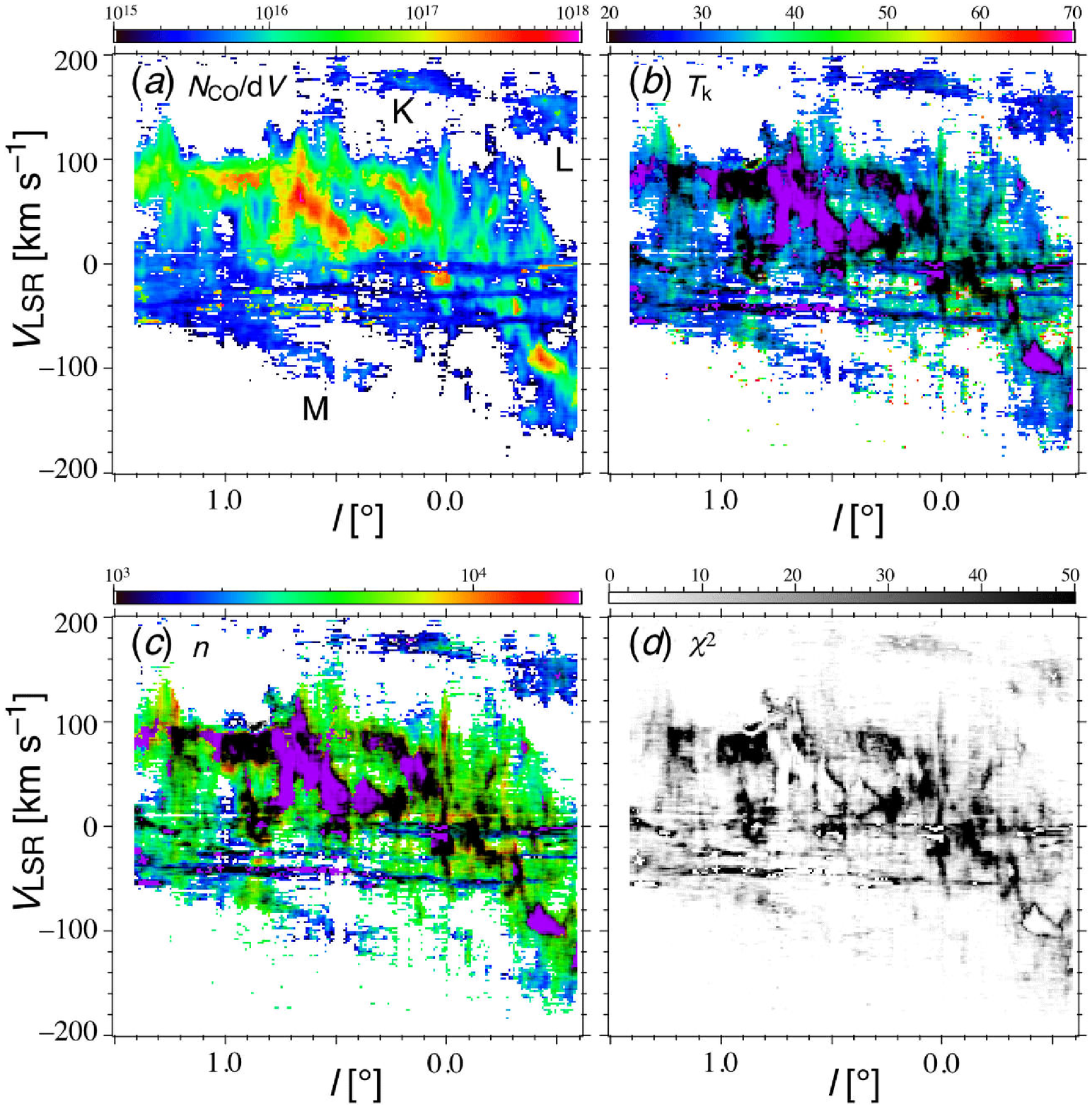}
  \end{center}
  \caption{Longitude-velocity diagrams at $b$=$-0.009\arcdeg$ of the physical conditions (color); ($a$) CO column density per unit velocity dispersion interval, ($b$) kinetic temperature, ($c$) number density of molecular hydrogen, and ($d$) chi-square function. The diagram of chi-square function (gray) and areas of improper solutions (purple) are overlaid in ($b$) and ($c$). The structures K, L, and M \citep{2006A&A...455..963R} are indicated in ($a$). }\label{fig:lv}
\end{figure*}

Figure \ref{fig:lv} shows longitude-velocity diagrams of physical conditions at $b$=$-0.009\arcdeg$ for example. 
Data in the velocity range $-60$ km s$^{-1} < V_{\rm LSR} < 20$ km s$^{-1}$, are severely contaminated by the four spiral arms in the Galactic disk. 
However, since our analyses are inapplicable to opaque gas, we exclude this velocity range from the following discussion. 
For the data points in the velocity ranges, $V_{\rm LSR} < -60$ km s$^{-1}$ and 20 km s$^{-1}$ $< V_{\rm LSR}$, we have proper LVG solutions for 281885 data points out of 410704 data points (68.6 \% ). 
Here we describe physical conditions of the two major components in the CMZ: the 120-pc star forming ring \citep{1995PASJ...47..527S} and the outer dust lanes. 

\subsubsection{Star-Forming Ring}
A large fraction of dense molecular gas in the CMZ is confined in the 120-pc star-forming ring, which contains Sgr B1, B2 and Sgr C HII regions. 
Our calculations show that the star-forming ring typically has kinetic temperature of 20 $\sim$ 35 K and density of 10$^{3.5-4.0}$ cm$^{-3}$. 
There are some areas where our LVG model cannot reproduce its line intensities, being associated with spots of improper solutions. In these areas, the opaque CO $J$=1--0 line is likely self-absorbed by low-density envelope of molecular clouds. 

The result is consistent with previous works of NH$_3$ lines \citep{1993A&A...280..255H} and those of high density tracers (e.g. \cite{1999ApJS..120....1T}). The low-$J$ CO lines arise from the cool, dense component suggested by \citet{1993A&A...280..255H}. \citet{2004ApJS..150..239M} reported the results of LVG analysis using [CI], CO $J$=7--6 and $J$=4--3 lines. They found kinetic temperatures about 50 K; this is higher than ours, and the discrepancy may be due to the difference in transitions used.

\subsubsection{Outer dust lanes}
The structures K, L, and M \citep{2006A&A...455..963R}, which had been recognized as parts of the 200-pc expanding molecular ring \citep{1972Natur.238..105K}, are now understood as dust lanes associated with the central bar. 
They have roughly uniform kinetic temperature about 30 K and low density 10$^{3.0-3.5}$ cm$^{-3}$. 
The density is consistent with the non-detections of high-density tracers. 
Hot (250 K) and low-density ($ \leq 10^{2}$ cm$^{-3}$) component associated with the structure M has been detected by the H$_3^+$ absorption line observations toward the Quintuplet cluster \citep{2005ApJ...632..882O}. However, we found no strong evidence of high temperature in the structure M. 
Two explanations for the disagreement may be possible; (1) hot gas detected in H$_3^+$ absorption has extremely low density which cannot excite the $J$=1 level of CO; (2) hot gas is spatially localized to the Quintuplet cluster.

\subsection{Mass Estimate}
\begin{figure*}[btp]
  \begin{center}
    \FigureFile(160mm,80mm){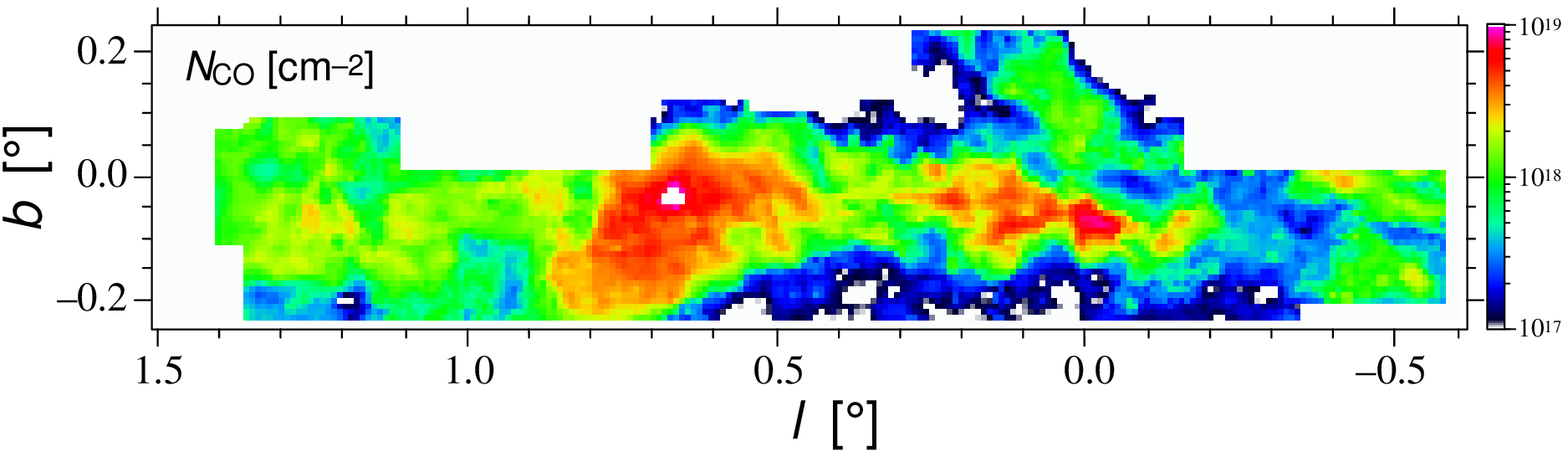}
  \end{center}
  \caption{A map of CO column density for the velocity ranges from $V_{\rm LSR}=-200$ km s$^{-1}$ to $-60$ km s$^{-1}$ and from $+20$ km s$^{-1}$ to $+200$ km s$^{-1}$.}\label{fig:column}
\end{figure*}

We made a map of CO column density in the CMZ by summing up $N_{\rm CO}$/d$V$ over the velocity ranges from $V_{\rm LSR}=-200$ km s$^{-1}$ to $-60$ km s$^{-1}$ and from $+20$ km s$^{-1}$ to $+200$ km s$^{-1}$ (Fig. \ref{fig:column}). 
The maximum $N_{\rm CO}$ is 3.1$ \times 10^{19}$ cm$^{-2}$ found in the direction of the Sgr B2 core. 
This value is consistent with the $^{13}$CO column density 9$ \times 10^{17}$ cm$^{-2}$ \citep{1989ApJ...337..704L}. 
Assuming that [CO]/[H$_2$] is 2.4$ \times 10^{-5}$, based on [$^{13}$CO]/[H$_2$] given by \citet{1989ApJ...337..704L}, molecular mass included in our data is $ \sim 2 \times 10^{7}$ $M_{\odot}$. This amount is smaller than the known value of molecular gas mass in the CMZ, 5--10$ \times 10^{7}$ $M_{\odot}$ \citep{1996ARA&A..34..645M}. The discrepancy may be accounted by; (1) incomplete coverage of our data for the entire CMZ, (2) inapplicability of our analyses to high $N_{\rm CO}/{\rm d}V$ regions.

\subsection{High CO $J$=3--2/$J$=1--0 Intensity Ratio Gas}
\begin{figure*}[tbp]
  \begin{center}
    \FigureFile(160mm,80mm){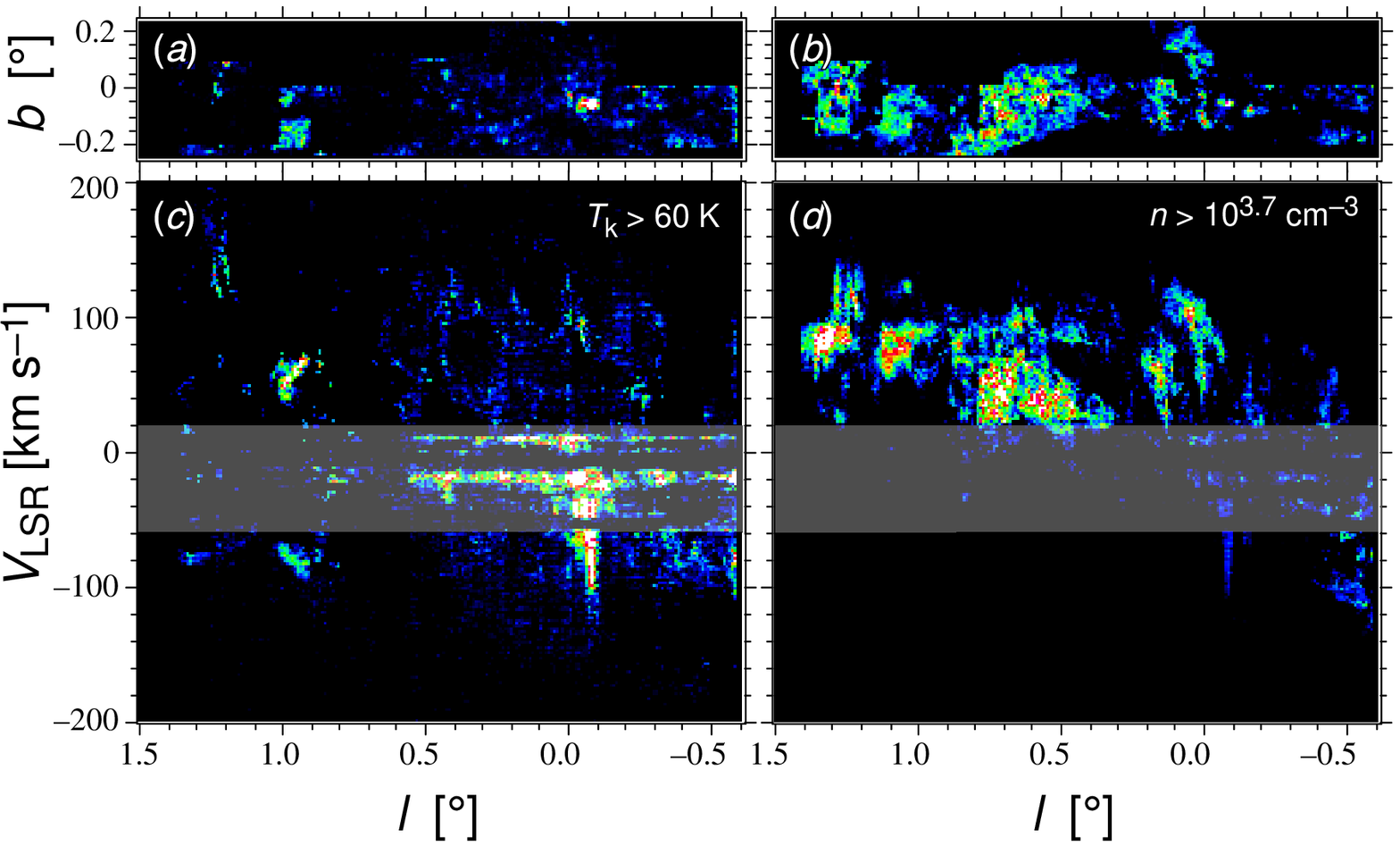}
  \end{center}
  \caption{({\it a}) A map of CO $J$=3--2 emission ($ \int T_{3-2}{\rm d}V$) integrated over velocity ranges from $V_{\rm LSR}=-200$ km s$^{-1}$ to $-60$ km s$^{-1}$ and from $+20$ km s$^{-1}$ to $+200$ km s$^{-1}$ for data with kinetic temperatures higher than 60 K. ({\it b}) Longitude-velocity diagrams of CO $J$=3--2 emission integrated over available latitudes ($ \sum T_{3-2}$)for data with kinetic temperatures higher than 60 K. ({\it c}) and ({\it d}) Same as ({\it a}) and ({\it b}) respectively but with densities higher than $10^{3.7}$ cm$^{-3}$. }\label{fig:hthd}
\end{figure*}

The CO $J$=3--2/$J$=1--0 intensity ratio, $R_{3-2/1-0} \equiv T_{3-2}/T_{1-0}$ may be useful to diagnose the physical conditions. 
Generally, high $R_{3-2/1-0}$ exceeding 1.5 is a signature of highly-excited, optically thin gas with high temperature and high density. 
Figure \ref{fig:hthd} shows the CO $J$=1--0 distribution of high temperature gas, $T_{\rm k} \geq 60$ K, and high density gas $n \geq 10^{3.7}$ cm$^{-3}$. 
The distribution of high $R_{3-2/1-0}$ gas (Fig. 5 of \cite{paper1}) is strikingly similar to that of high temperature gas. 
This suggests that $R_{3-2/1-0}$ is a good measure of kinetic temperature, demonstrating validity of extracting shocked gas by high $R_{3-2/1-0}$. 

\section{Summary}
We developed a robust scheme to estimate physical conditions by fitting results of LVG calculations to multi-line data sets.  We applied the scheme to the CO {\it J}=3--2 data of the Galactic center CMZ obtained with the ASTE and the CO, $^{13}$CO {\it J}=1--0 data sets in the same region obtained with the NRO 45 m telescope.  The analyses made a success to some degree, properly determining physical conditions within 99\%\
 confidence level for about 69 \%\
 of the data points.  
The physical conditions in the CMZ are summarized as follows; 
\begin{itemize}
\item the 120-pc star forming ring has higher density than the outer dust lanes. 
\item the kinetic temperature of the outer dust lanes are uniformly low, $\sim 30$ K, 
\item high $R_{3-2/1-0}$ ($ \geq 1.5$) gas in this region is mostly high temperature gas.  
\end{itemize}

We could not determine physical conditions in high-density cores with large column density, since we used two opaque lines with rather low effective critical densities.  Breakdowns of the one-zone assumption might also worsen the accuracy of our analysis.  The use of less opaque lines such as C$^{18}$O, as well as high-density tracers must improve the accuracy.  \\


The authors would like to thank Prof. T. Hasegawa for helpful comments. M. N. is indebted to his thesis advisor Prof. S. Yamamoto for his support. 
A part of this study was financially 
supported by the MEXT Grant-in-Aid for Scientific 
Research on Priority Areas No. 15071202.



\end{document}